\documentclass[12pt]{article}

\usepackage{axodraw}
\usepackage{epsf}
\usepackage{rotating}

\catcode`@=11
\def\citer{\@ifnextchar
[{\@tempswatrue\@citexr}{\@tempswafalse\@citexr[]}}

\def\@citexr[#1]#2{\if@filesw\immediate\write\@auxout{\string\citation{#2}}\fi
  \def\@citea{}\@cite{\@for\@citeb:=#2\do
    {\@citea\def\@citea{--\penalty\@m}\@ifundefined
       {b@\@citeb}{{\bf ?}\@warning
       {Citation `\@citeb' on page \thepage \space undefined}}%
\hbox{\csname b@\@citeb\endcsname}}}{#1}}
\catcode`@=12

\newcommand{\beq}{\begin{eqnarray}}
\newcommand{\eeq}{\end{eqnarray}}
\newcommand{\nn}{\noindent}

\newcommand{\tgb}{\tan\beta}

\newcommand{\st}{\tilde{t}}

\newcommand{\sq}{\tilde{q}}
\newcommand{\sqb}{\bar{\tilde{q}}}
\newcommand{\gl}{\tilde{g}}

\newcommand{\MSSM}{\mbox{$\MSSM}$}

\newcommand{\npb}[3]{${\rm Nucl.~Phys.}$ {\bf B#1} (19#2)~#3}
\newcommand{\plb}[3]{${\rm Phys.~Lett.}$ {\bf B#1} (19#2) #3}

 \newcommand{\zp}[3]{{Z.\ Phys.} {\bf #1} (19#2) #3}
 \newcommand{\np}[3]{{Nucl.\ Phys.} {\bf #1} (19#2)~#3}
 \newcommand{\pl}[3]{{Phys.\ Lett.} {\bf #1} (19#2) #3}
 \newcommand{\pr}[3]{{Phys.\ Rev.} {\bf #1} (19#2) #3}
 \newcommand{\prl}[3]{{Phys.\ Rev. Lett.} {\bf #1} (19#2) #3}

\def\sla#1{\ifmmode%
\setbox0=\hbox{$#1$}%
\setbox1=\hbox to\wd0{\hss$/$\hss}\else%
\setbox0=\hbox{#1}%
\setbox1=\hbox to\wd0{\hss/\hss}\fi%
#1\hskip-\wd0\box1 } 

\newcommand{\lsim}{\raisebox{-0.13cm}{~\shortstack{$<$ \\[-0.07cm] $\sim$}}~}
\newcommand{\gsim}{\raisebox{-0.13cm}{~\shortstack{$>$ \\[-0.07cm] $\sim$}}~}
\newcommand{\fbi}{~fb$^{-1}\;$}
\begin{document}
 
\renewcommand{\thefootnote}{\fnsymbol{footnote} }

\begin{flushright}
PSI--PR--02--19 \\
hep-ph/0211145
\end{flushright}

\begin{center}

{\large\sc Higgs and SUSY Particle Production at Hadron Colliders}%
\footnote{Talk presented at SUSY02, DESY Hamburg, 2002. \\
This work has been supported in part by the Swiss Bundesamt f\"ur Bildung
und Wissenschaft and by the European Union under contract
HPRN-CT-2000-00149.}

\end{center}

\begin{center}
Michael Spira

\vspace*{0.3cm}

{\it \small Paul Scherrer Institut, CH-5232 Villigen PSI, Switzerland}
\end{center}
\vspace*{0.2cm}

\begin{abstract}
\nn
The theoretical status of Higgs boson and supersymmetric particle
production at hadron colliders is reviewed with particular emphasis on
recent results and open problems.
\end{abstract}

\section{Introduction}
Supersymmetry imposes a new symmetry between the fermionic and bosonic
degrees of freedom \cite{susy}. Since supersymmetric theories do not
develop quadratic divergences in higher orders, they provide a natural
solution of the hierarchy problem at the electroweak scale
\cite{hierarchy}. If supersymmetric grand unified theories (SUSY-GUT) are
considered, the predicted value of the Weinberg angle turns out to be in
excellent agreement with the present measurements at the LEP and SLC
experiments \cite{gutsw}. Moreover, owing to the large top quark mass
SUSY-GUTs develop electroweak symmetry breaking at the electroweak scale
dynamically \cite{esbdyn}. Due to these properties the supersymmetric
extension of the Standard Model exhibits one of the most attractive
alternatives beyond the Standard Model.

The Higgs mechanism is a cornerstone of the Standard Model (SM) and its
supersymmetric extensions \cite{higgs}. Thus, the search for Higgs bosons
is one of the most important endeavors at future high-energy experiments.
The minimal supersymmetric extension of the Standard Model (MSSM) requires
the introduction of two Higgs doublets in order to preserve supersymmetry.
There are five elementary Higgs particles, two CP-even ($h,H$), one CP-odd
($A$) and two charged ones ($H^\pm$). At lowest order all couplings and
masses of the MSSM Higgs sector are fixed by two independent input
parameters, which are generally chosen as $\tgb=v_2/v_1$, the ratio of the
two vacuum expectation values $v_{1,2}$, and the pseudoscalar Higgs-boson
mass $M_A$. At LO the light scalar Higgs mass $M_h$ has to be smaller than
the $Z$-boson mass $M_Z$. Including the one-loop and dominant two-loop
corrections the upper bound is increased to $M_h\lsim 135$ GeV
\cite{mssmrad}. The negative direct searches for the Higgsstrahlung
processes $e^+e^-\to Zh,ZH$ and the associated production $e^+e^-\to
Ah,AH$ yield lower bounds of $m_{h,H} > 91.0$ GeV and $m_A > 91.9$ GeV.
The range $0.5 < \tgb < 2.4$ in the MSSM is excluded by the Higgs searches
at the LEP2 experiments \cite{lep2}.

Higgs bosons can be searched for at the upgraded Tevatron, a $p\bar p$
collider with a c.m.\ energy of 2 TeV, and the LHC, a $pp$ collider with a
c.m.\ energy of 14 TeV. At the Tevatron the most important processes are
Higgs-strahlung $q\bar q \to W+h/H$ with $h/H\to b\bar b$ which is
important for Higgs masses below about 130 GeV, gluon fusion $gg\to h/H\to
W^* W$ which is relevant for Higgs masses above 130 GeV, and Higgs
radiation off bottom quarks $q\bar q\to b\bar b\phi$ which plays an
important r\^ole for large values of $\tgb$. With an integrated luminosity
of 30\fbi the Tevatron can probe the entire MSSM parameter space
\cite{run2}. At the LHC the most important Higgs production modes are
gluon fusion, vector boson fusion $qq\to qq+h/H$ and Higgs radiation off
top and bottom quarks. There are several Higgs decay modes which enable
the discovery of the Higgs bosons \cite{lhc}.

The novel colored particles, squarks and gluinos, and the weakly
interacting gauginos can be searched for at the Tevatron and the LHC.
Until now the search at the Tevatron has set the most stringent bounds on
the colored SUSY particle masses.  At the 95\% CL, gluinos have to be
heavier than about 180 GeV, while squarks with masses below about 180 GeV
have been excluded for gluino masses below $\sim 300$ GeV \cite{bounds}.  
Stops, the scalar superpartners of the top quark, have been excluded in a
significant part of the MSSM parameter space with mass less than about
80--100 GeV by the LEP \cite{lep} and Tevatron experiments \cite{bounds}.  
Finally charginos with masses below about 100 GeV have been excluded by the
LEP experiments \cite{lep}, while the present search at the Tevatron is
sensitive to chargino masses of about 60--80 GeV with a strong dependence
on the specific model \cite{trilepton}. Due to the negative search at LEP2
the lightest neutralino $\tilde \chi_1^0$ has to be heavier than about 45
GeV in the context of SUGRA models \cite{lep}. In the
$R$-parity-conserving MSSM, supersymmetric particles can only be produced
in pairs.  All supersymmetric particles will decay to the lightest
supersymmetric particle (LSP), which is most likely to be a neutralino,
stable thanks to conserved $R$-parity.  Thus the final signatures for the
production of supersymmetric particles will mainly be jets, charged
leptons and missing transverse energy, which is carried away by neutrinos
and the invisible neutral LSP.

\section{Higgs boson production}
\subsection{Gluon fusion}
The gluon fusion mechanism $gg\to \phi$ provides the dominant production
mechanism of Higgs bosons at the LHC in the entire relevant mass range up
to about 1 TeV for small and moderate values of $\tgb$ in the MSSM
\cite{habil}. At the Tevatron this process plays a r\^ole for Higgs masses
between about 130 GeV and 190 GeV, if the branching ratio of decays
into $W^*W$ pairs is large enough \cite{run2}. The gluon fusion process is
mediated by heavy quark triangle loops and, in the case of supersymmetric
theories, by squark loops in addition, if the squark masses are smaller
than about 400 GeV \cite{squark}.

\begin{figure}[hbt]
\vspace*{-0.8cm}
\hspace*{4.0cm} \includegraphics[width=8cm]{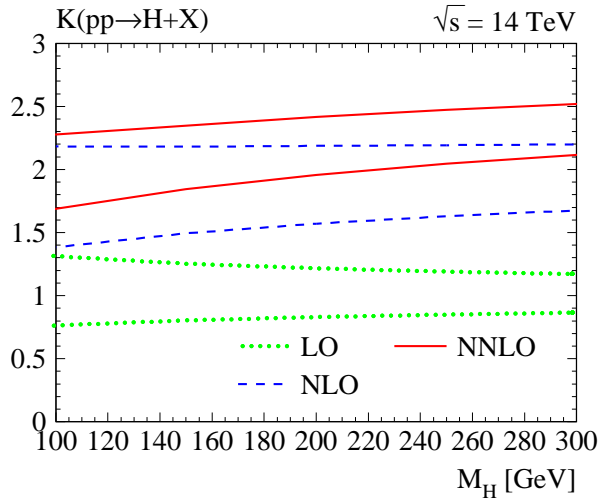}
\vspace*{-0.4cm}
      \caption[]{\it \label{fig:all14murf} Scale dependence of the $K$-factor 
      at the LHC. Lower curves for each pair are for 
      $\mu_R = 2M_H$, $\mu_F=M_H/2$, upper curves are for 
      $\mu_R =M_H/2$, $\mu_F=2M_H$.  The $K$-factor is
      computed with respect to the LO cross section at
      $\mu_R = \mu_F =M_H$. From Ref.~\cite{glufusnnlo}.}
\end{figure}

In the past the full two-loop QCD corrections have been determined. They
increase the production cross sections by 10--90\%
\cite{glufusnlo0,glufusnlo}, thus leading to a significant change of the
theoretical predictions.  Very recently, the full NNLO calculation has
been finished in the heavy top quark limit~\cite{glufusnnlo}. This limit
has been demonstrated to approximate the full massive $K$ factor at NLO
within about 20\% for small $\tgb$ in the entire mass range up to 1 TeV
\cite{softgluon}. Thus, a similar situation may be expected at NNLO. The
reason for the quality of this approximation is that the QCD corrections
to the gluon fusion mechanism are dominated by soft and collinear gluon
effects, which do not resolve the one-loop Higgs coupling to gluons.
Fig.~\ref{fig:all14murf} shows the resulting $K$-factors at the LHC and
the scale variation of the $K$-factor for the SM Higgs boson. The
calculation stabilizes at NNLO, with remaining scale variations at the
10--15\% level.  These uncertainties are comparable to the experimental
errors which can be achieved with 300\fbi of data at the LHC \cite{lhc}.
The full NNLO results confirm earlier estimates which were obtained in the
frame work of soft gluon resummation \cite{softgluon} and soft
approximations \cite{soft} of the full three-loop result within 10--15\%.
The full soft gluon resummation has been performed in Ref.\,\cite{soft2}.
The resummation effects enhance the NNLO result further by about 10\% thus
signaling a perturbative stabilization of the theoretical prediction for
the gluon-fusion cross section.

In supersymmetric theories the gluon fusion cross sections for the heavy
Higgs, $H$, and, for small $M_A$, also for the light Higgs, $h$, are
significantly affected by bottom quark loops for $\tgb\gsim 3$ so that the
heavy top quark limit is not applicable in general. This can be clearly
seen in the NLO results, which show a decrease of the $K$ factor down to
about 1.1 for large $\tgb$ \cite{glufusnlo}. This decrease
originates from an interplay between the large positive soft/collinear
gluon effects and large negative double logarithms of the ratio between
the Higgs and bottom masses.  In addition, the shape of the $p_T$
distribution of the Higgs boson may be altered; if the bottom loop is
dominant, the $p_T$ spectrum becomes softer than in the case of top-loop
dominance. These effects lead to some model dependence of predicted cross
sections.

\subsection{$t\bar t\phi$ production}
\begin{figure}[hbt]
\vspace*{-2.0cm}
\hspace*{0.0cm}
\epsfxsize=7.5cm \epsfbox{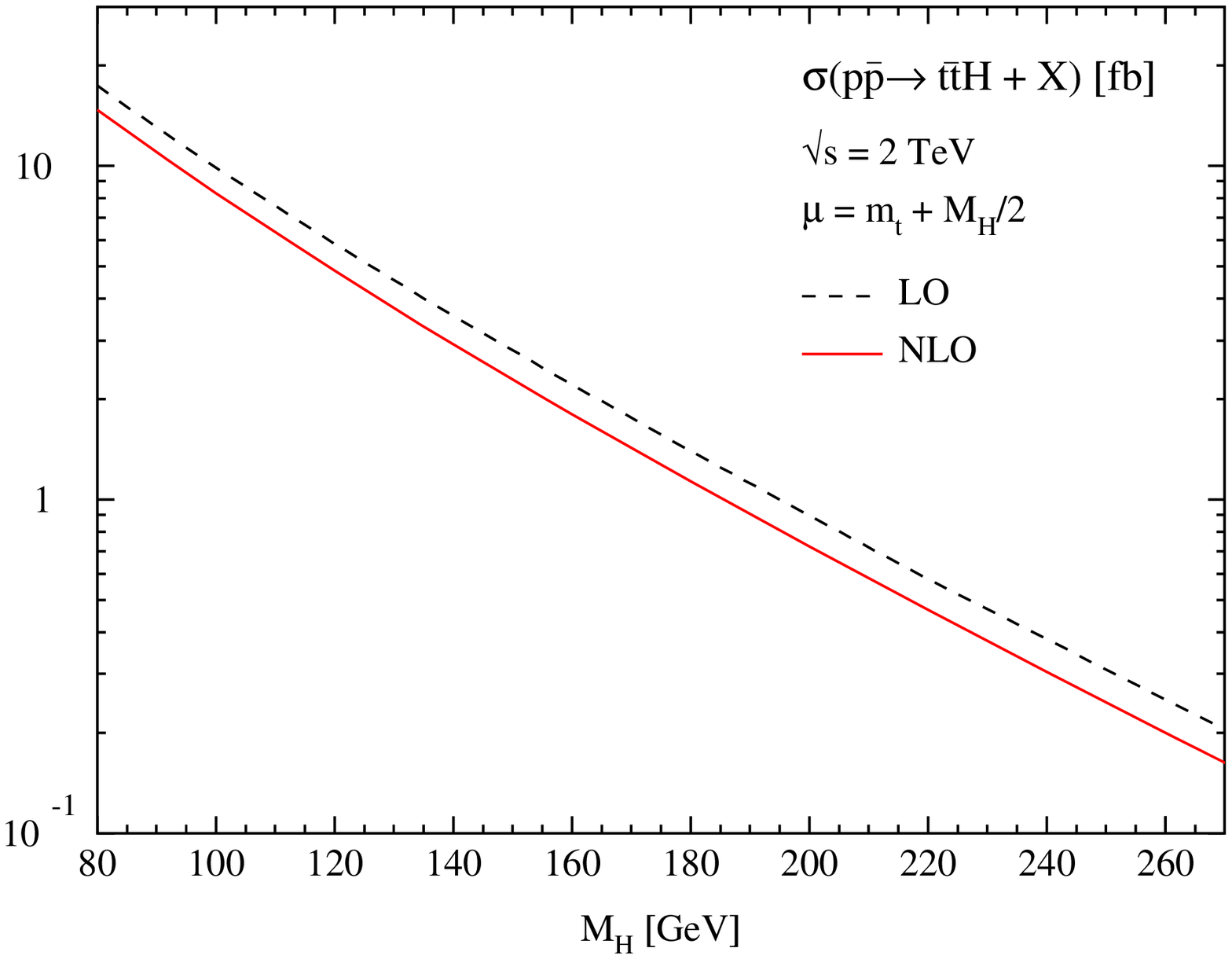}
\vspace*{-10.40cm}

\hspace*{8.3cm}
\epsfxsize=7.5cm \epsfbox{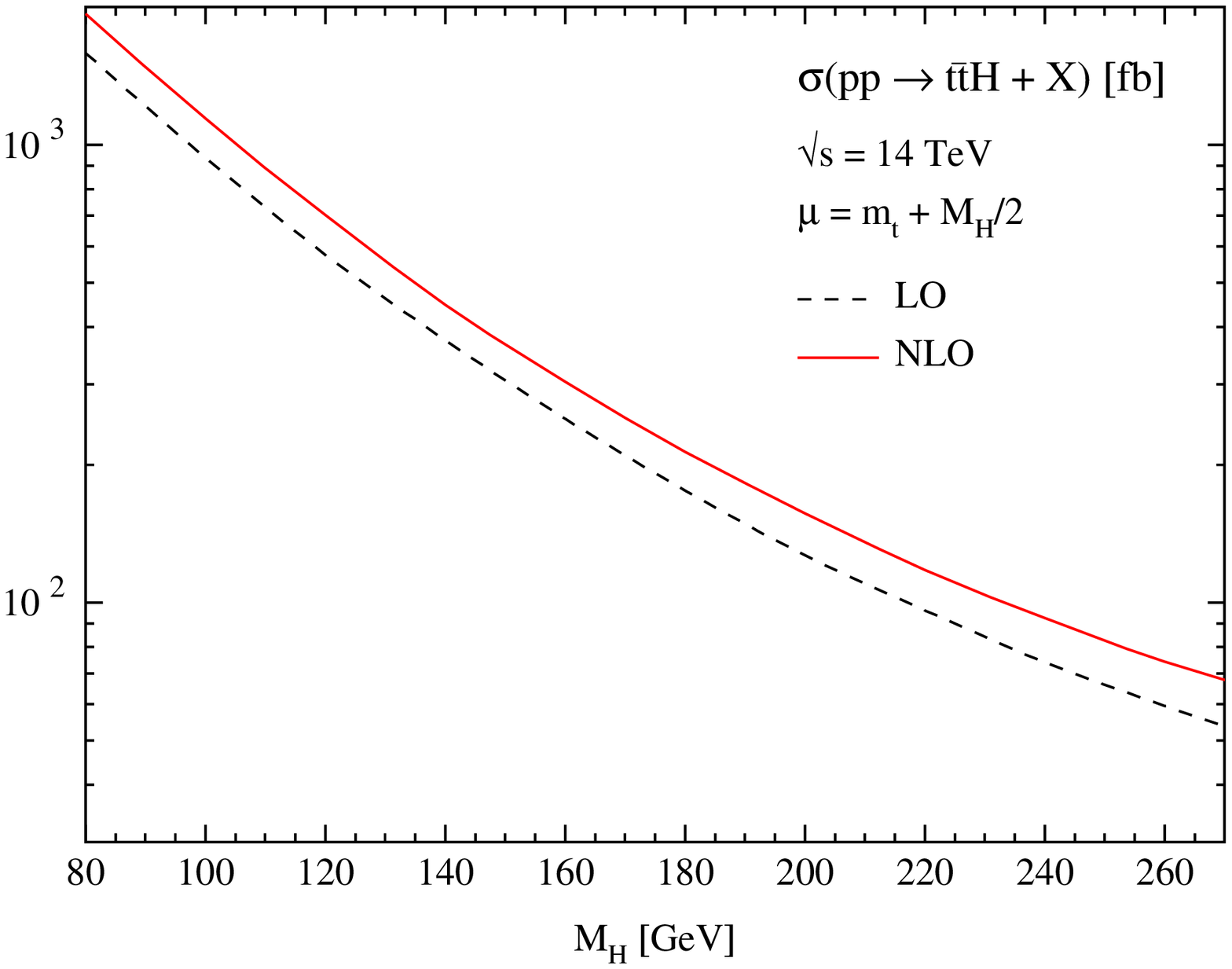}
\vspace*{-2.8cm}

\caption[]{\label{fg:tth} \it The cross section for
$pp/p\bar p\to t\bar tH\;+\;X$ at the Tevatron and
LHC in LO and NLO approximation, with the renormalization and
factorization scales set to $\mu=m_t+M_H/2$.}
\end{figure}

SM Higgs boson production in association with $t\bar t$ pairs plays a
significant r\^ole at the LHC for Higgs masses below about 130 GeV, since
this production mechanism makes the observation of $H\to b\bar b$ possible
\cite{lhc,drollinger}.  The decay
$H\to\gamma\gamma$ is potentially visible in this channel at high
integrated luminosity. For Higgs masses above about 130 GeV, the decay
$H\to W^*W$ can be observed~\cite{tth2ww}.  $t\bar tH$ production could
conceivably be used to determine the top Yukawa coupling directly from the
cross section. NLO QCD
corrections have become available. They decrease the cross section at the
Tevatron by about 20\% \cite{tthnlo,tthnloq}, while they increase the
signal rate at the LHC by about 20--40\% \cite{tthnlo}, see
Fig.~\ref{fg:tth}. The scale dependence of the production cross section is
significantly reduced, to a level of about 10--15\%, which can be
considered as an estimate of the theoretical uncertainty. The transverse
momentum and rapidity distributions at NLO can be approximated by a
rescaling of the LO distributions with a constant $K$ factor within
10--15\% \cite{tthnlolong}. Thus, the signal rate is under proper
theoretical control now. In the MSSM, $t\bar th$ production with $h\to
\gamma\gamma,b\bar b$ is important at the LHC in the decoupling regime,
where the light scalar $h$ behaves as the SM Higgs boson
\cite{lhc,drollinger}.  Thus, the SM
results can also be used for $t\bar{t}h$ production in this regime.

\subsection{$b\bar b\phi$ production}
In supersymmetric theories $b\bar b\phi$ production becomes the dominant
Higgs boson production mechanism for large values of $\tgb$ \cite{habil},
where the bottom Yukawa coupling is strongly enhanced. In contrast to
$t\bar t\phi$ production, however, this process develops potentially large
logarithms, $\log m_\phi^2/m_b^2$, in the high-energy limit due to the
smallness of the bottom quark mass, which are related to the development
of $b$ densities in the initial state. They can be resummed by evolving
the $b$ densities according to the DGLAP--equations and introducing them
in the production process \cite{willenbrock}. The NLO QCD corrections to
the $b$-initiated processes $b\bar b\to H$ \cite{bb2h} and
$\raisebox{-0.00cm}{\shortstack{{\tiny (---)} \\[-0.15cm] $b$}} g \to
\raisebox{-0.00cm}{\shortstack{{\tiny (---)} \\[-0.15cm] $b$}} H$
\cite{bg2bh} are known to be moderate. The resummation increases the
cross section by a factor of about 5 at the Tevatron and about 2--3 at
the LHC and thus plays a significant phenomenological r\^ole. The
introduction of conventional $b$ densities, however, requires an
approximation of the hard process kinematics, i.e.~the initial and final
$b$ quarks are assumed to be massless and travel predominantly in forward
and backward
direction. These approximations can be tested in the full $gg\to b\bar
b\phi$ process.

We have to investigate if the energy of the
Tevatron and LHC is sufficiently large to develop the factorization of
bottom densities, i.e.~that the transverse mass distribution of the
(anti)bottom quark can be factorized as a convolution
\beq
\frac{d\sigma}{dM_{Tb}} = \frac{1}{M_{Tb}} \left\{ \frac{\alpha_s}{2\pi}
P_{qg} \otimes g \otimes g \otimes \hat\sigma_{bg} \right\}_{M_{Tb} = m_b
\to 0} + \mbox{non-singular terms}
\label{eq:bfact}
\eeq
where $M_{Tb}=\sqrt{m_b^2+p_{Tb}^2}$ denotes the transverse mass of the
(anti)bottom quark, $P_{qg}$ the corresponding DGLAP splitting kernel,
$g$ the gluon density of the (anti)proton and $\hat \sigma_{bg}$ the
partonic cross section for $\raisebox{-0.00cm}{\shortstack{{\tiny (---)}
\\[-0.15cm] $b$}} g \to \raisebox{-0.00cm}{\shortstack{{\tiny (---)}
\\[-0.15cm] $b$}} H$. This factorization requires
that the transverse mass distribution is dominated by the first term,
i.e.~$d\sigma/dM_{Tb}\propto 1/M_{Tb}$, for transverse masses up to the
factorization scale of the (anti)bottom density.
\begin{figure}[hbt]
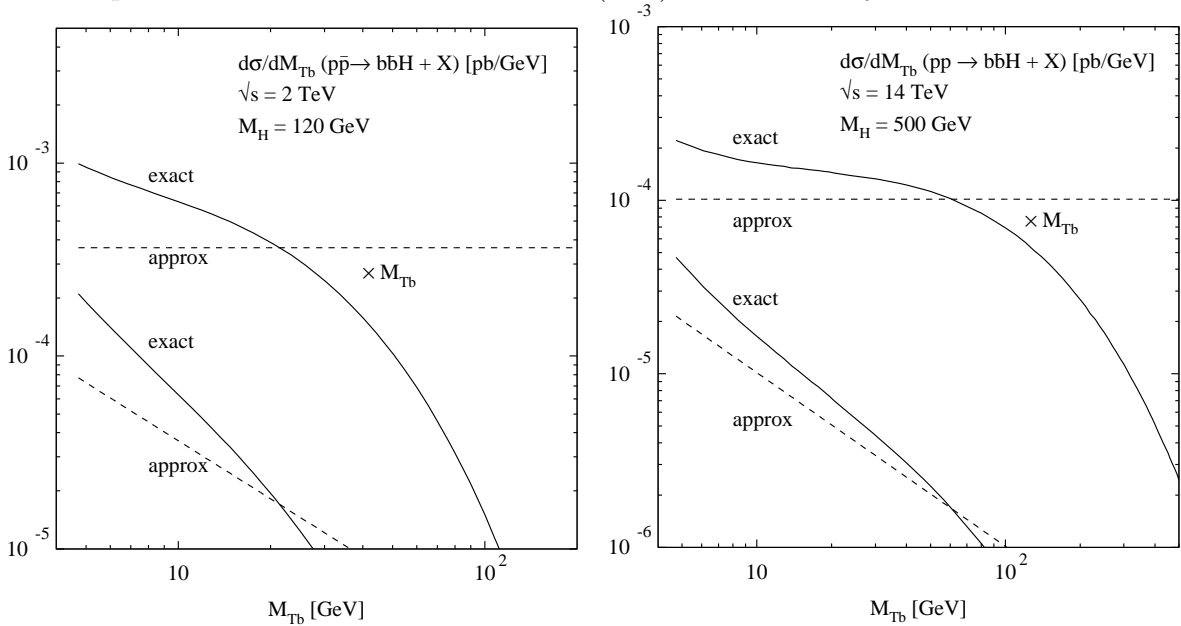

\vspace*{-0.8cm}
\hspace*{0.5cm}
\epsfxsize=7cm \epsfbox{pttev.120}
\vspace*{-9.75cm}

\hspace*{8.5cm}
\epsfxsize=7cm \epsfbox{ptlhc.500}
\vspace*{-1.5cm}

\caption[]{\label{fg:bbh} \it Transverse mass distributions of the bottom
quark in $b\bar bH$ production at the Tevatron and the LHC. We have
adopted CTEQ5M1 parton densities and a bottom mass of $m_b=4.62$ GeV. The
solid lines show the full LO result from $q\bar q,gg\to b\bar bH$ and the
dashed lines the factorized collinear part of Eq.~(\ref{eq:bfact}) which
is absorbed in the bottom parton density. The upper curves are multiplied
with the factor $M_{Tb}$ of the asymptotic behavior, which is required by
factorizing bottom densities.}
\end{figure} 
The transverse mass distributions at the Tevatron and LHC are shown in
Fig.~\ref{fg:bbh}. The solid curves show the full distributions of the
$q\bar q,gg\to b\bar b\phi$ processes, while the dashed lines exhibit the
factorized collinear part of Eq.~(\ref{eq:bfact}) which is absorbed in the
bottom density. For a proper factorization, these pairs of curves have to
coincide approximately up to transverse masses of the order of the
factorization scale, which is usually chosen to be $\mu_F={\cal O}(m_H)$.
It is clearly visible that there are sizeable differences between the full
result and the factorized part, which originate from sizeable bottom mass
and phase space effects, that are not accounted for by an active bottom
parton density. Moreover, the full result falls quickly below the
approximate factorized part for transverse masses of the order of
$m_H/10$, which is much smaller than the usual factorization scale used
for the bottom densities. We conclude from these plots that $b\bar b\phi$
production at the Tevatron and LHC develops sizeable bottom mass and
kinematical phase
space effects, so that the use of bottom densities in the process $b\bar
b\to \phi$ may lead to an overestimate of the correct theoretical result
due to too crude approximations in the kinematics of the hard process
\cite{houches}. The
full NLO calculation of the $gg\to b\bar b\phi$ will yield much more
insight into this problem, since the large logarithms related to the
evolution of bottom densities have to appear in the NLO corrections, if
the picture of active bottom quarks in the proton is correct.

\section{SUSY particle production}
\subsection{Production of squarks and gluinos}
Squarks and gluinos can be produced via $pp, p\bar p \to \sq \sqb, \sq
\sq, \sq \gl, \gl \gl$ at hadron colliders. The determination of the full
SUSY--QCD corrections has been performed for the upgraded Tevatron and the
LHC. For the natural renormalization/factorization scale choice $Q=m$,
where $m$ denotes the average mass of the final-state SUSY particles, the
SUSY QCD corrections are large and positive, increasing the total cross
sections by 10--90\% \cite{sqgl}. The inclusion of the NLO corrections
reduces the LO scale dependence by a factor 3--4 and reaches a typical
level of $\sim$ 10--15\% which serves as an estimate of the remaining
theoretical uncertainty.  Moreover, the dependence on different sets of
parton densities is rather weak and leads to an additional uncertainty of
$\sim$ 10--15\%. In order to quantify the effect of the NLO corrections on
the search for squarks and gluinos at hadron colliders, the SUSY particle
masses corresponding to several fixed values of the production cross
sections have been extracted. These masses are increased by 10--30 GeV at
the Tevatron and 10--50 GeV at the LHC, thus enhancing the present and
future bounds on the squark and gluino masses significantly.
Finally, the QCD-corrected transverse-momentum and rapidity distributions
for all different processes have been evaluated. The modification of the
normalized distributions in NLO compared to LO is less than about 15\% for
the transverse-momentum distributions and much less for the rapidity
distributions.  Thus it is a sufficient approximation to rescale the LO
distributions uniformly by the K factors of the total cross sections
\cite{sqgl}.

\subsection{Stop pair production}
At LO only pairs of $\st_1$ or pairs of $\st_2$ can be produced at hadron
colliders. QCD-initiated mixed $\st_1 \st_2$ pair production is only
possible at NLO
and beyond.  However, mixed stop pair production is completely suppressed
by several orders of magnitude and can thus safely be neglected
\cite{stops}. The evaluation of the SUSY--QCD corrections proceeds along
the same lines as in the case of squarks and gluinos. They increase the
total cross sections by up to about 40\% \cite{stops}. As in the
squark/gluino case the scale dependence is strongly reduced and yields an
estimate of about 10--15\% of the remaining theoretical uncertainty at
NLO. At NLO the virtual corrections depend on the stop mixing angle, the
squark, gluino and stop masses of the other type. However, it turns out
that these dependences are very weak and can safely be neglected.

\subsection{Chargino and neutralino production}
The production cross sections of charginos and neutralinos depend on
several MSSM parameters, i.e.\ $M_1, M_2, \mu$ and $\tgb$ at LO \cite{lo}.
The cross sections are sizeable for chargino/neutralino masses below about
100 GeV at the upgraded Tevatron and less than about 200 GeV at the LHC.
Due to the strong dependence on the MSSM parameters the extracted bounds
on the chargino and neutralino masses depend on the specific region in the
MSSM parameter space \cite{bounds}. The outline of the determination of
the SUSY--QCD corrections is analogous to the previous cases of squarks,
gluinos and stops. The corrections enhance the production cross sections of
charginos and neutralinos by about 10--40\% \cite{gaunlo}. The scale
dependence is reduced to about 10\% at NLO, which signalizes a significant
stabilization of the theoretical prediction for the production cross
sections. The dependence of the chargino/neutralino production cross
sections on the specific set of parton densities ranges at about 10--15\%
\cite{gaunlo}.

\subsection{Associated production of gluinos and gauginos}
The cross sections of the associated gluino-gaugino production are
sizeable for the lightest chargino/neutralino states at the upgraded
Tevatron and the LHC if the gluino mass is less than about 400--500 GeV
\cite{lo}. The determination of the SUSY--QCD corrections is
analogous to the previous cases of squarks, gluinos, stops and gauginos.
The production cross sections are decreased by up to 3\% at the
Tevatron and increased by about 10--20\% at the LHC due to these corrections
\cite{gauglnlo} as can be inferred from
\begin{figure}[hbt]
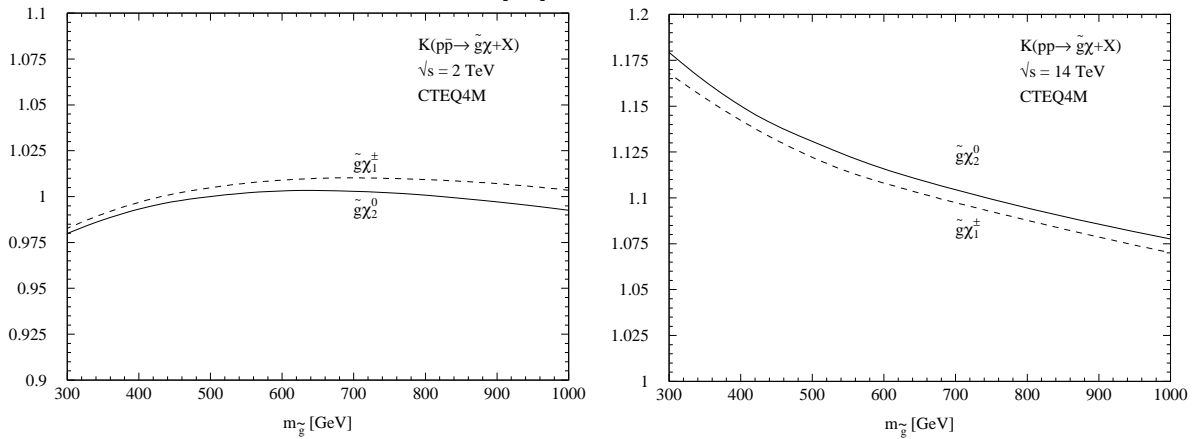

\vspace*{-0.2cm}
\hspace*{-0.3cm}
\begin{turn}{-90}%
\epsfxsize=5.5cm \epsfbox{gauglu.ktev}
\end{turn}
\vspace*{-5.52cm}

\hspace*{7.7cm}
\begin{turn}{-90}%
\epsfxsize=5.5cm \epsfbox{gauglu.klhc}
\end{turn}
\vspace*{-0.3cm}

\caption[]{\label{fg:gaugluk} \it K factor of the cross sections for
gluinos produced in association with the gauginos
$\chi^0_2$ and $\chi^\pm_1$ at the upgraded Tevatron
(left) and the LHC (right). Parton densities: CTEQ4L (LO) and CTEQ4M (NLO)
with the renormalization/factorization scale $Q=(m_{\gl}+m_\chi)/2$.}
\end{figure}
\begin{figure}[hbt]
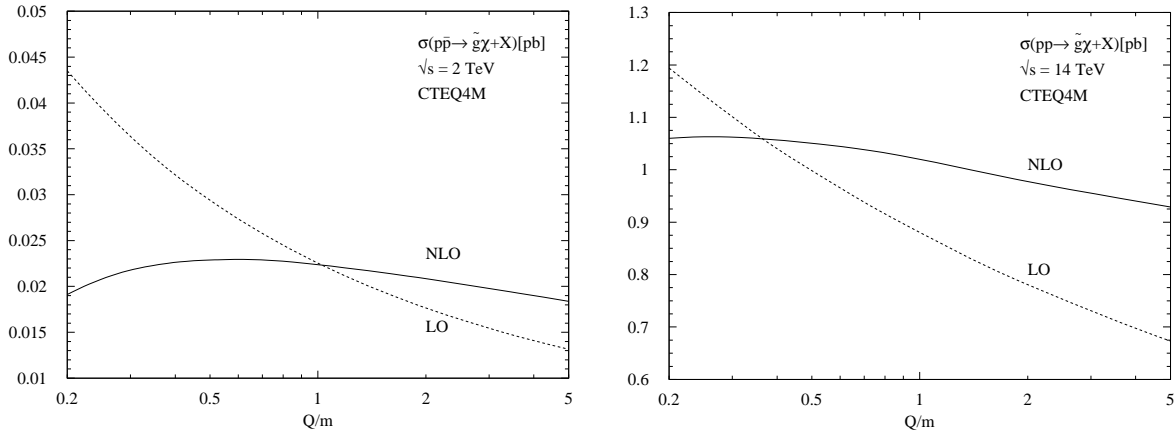

\vspace*{0.0cm}
\hspace*{-0.3cm}
\begin{turn}{-90}%
\epsfxsize=5.5cm \epsfbox{gauglu.sctev}
\end{turn}
\vspace*{-5.52cm}

\hspace*{7.7cm}
\begin{turn}{-90}%
\epsfxsize=5.5cm \epsfbox{gauglu.sclhc}
\end{turn}
\vspace*{-0.3cm}

\caption[]{\label{fg:gauglusc} \it Scale dependence of the cross sections of
gluinos produced in association with $\chi^0_2$ at the upgraded Tevatron
(left) and the LHC (right). Parton densities: CTEQ4L (LO) and CTEQ4M (NLO)
with the renormalization/factorization scale $Q$ varied in units of the
average mass $m=(m_{\gl}+m_\chi)/2$.}
\end{figure}
Fig.~\ref{fg:gaugluk}%
\footnote{These results disagree with the corresponding figures in
Ref.~\cite{berger}. The discrepancies have been resolved meanwhile and the
corrected calculations in the Erratum to Ref.~\cite{berger} are in agreement
with the results that we have presented to this conference. We are
grateful to the authors of Ref.~\cite{berger} for their cooperation.}.
The scale dependence is shown in Fig.~\ref{fg:gauglusc} at LO and NLO. It is
clearly visible that it is reduced to about 10--15\% at NLO which signalizes a
significant stabilization of the theoretical prediction for the production
cross sections. The dependence of the gluino-gaugino production cross
sections on the specific set of parton densities ranges at about 10--15\%
\cite{gauglnlo}.

\section{Conclusions}
Considerable progress has been made recently in improving QCD calculations
for Higgs and supersymmetric signal cross sections at hadron colliders.
Most (N)NLO [SUSY--]QCD corrections to all relevant production processes at
hadron colliders are known, i.e. the theoretical status of novel
particle production at the Tevatron and LHC is nearly complete. Large
corrections to many processes have been found which underlines the
importance of including them in realistic experimental analyses. After
inclusion of the NLO corrections the residual theoretical uncertainties
are reduced to a level of 10--15\%. There are several Fortran programs
which include most of the higher order corrections \cite{programs}.

\end{document}